\documentclass[12pt]{article}
\usepackage{amssymb}
\usepackage{epsfig}
\usepackage{graphicx}
\usepackage{amsmath}
\usepackage{verbatim}
\usepackage{titlesec}

\csname @addtoreset\endcsname{equation}{section}
\begin{document}
\begin{titlepage}
\title{\bf\Large  Is There Scale Invariance in $\mathcal{N}=1$ Supersymmetric Field Theories $?$\vspace{18pt}}
\author{\normalsize Sibo~Zheng \vspace{15pt}\\
{\it\small  Department of Physics, Chongqing University, Chongqing 401331, P.R. China}\\}

\date{}
\maketitle \voffset -.3in \vskip 1.cm \centerline{\bf Abstract}
\vskip .3cm In two dimensions, it is well known that the scale
invariance can be considered as conformal invariance.
However, there is no solid proof of this equivalence in four or higher dimensions.
we address this issue in the context of $4d$ $\mathcal{N}=1$ SUSY theories.
The SUSY version of dilatation current for theories without conserved $R$ symmetry is constructed through the FZ-multiplet.
 We discover that the scale-invariant SUSY theory is also conformal
 when the real superfield in the dilatation current multiplet is conserved.
 Otherwise, it is only scale-invariant, despite of the transformation of improvement.

 \vskip 5.cm
 \thispagestyle{empty}

\end{titlepage}
\newpage
\section{Introduction}
In two dimensions, the equivalence between scale and conformal invariance
has been proved in terms of a ``c theorem " \cite{Zamolodchikov, Polchinski}.
At the classical level this statement is argued to be still true in four-dimensional ( $4d$ ) QFT \cite{Callan}.
However, it is unclear at the quantum level generally,
although some examples ( e.g, see \cite{Polchinski, 0910.1087,0904.2715,9511154,9411149})
with or without supersymmetry (SUSY) have been proposed in the literature.

The difference between scale and conformal invariance in SUSY can be understood
from the viewpoint of structure of their groups \cite{Mack}.
The generators for the group of scale invariance include a dilatation operator $\Delta$ and those of
super-Poincare group.
The elements of the group for super-conformal symmetry are bigger.
In particular, the super-conformal group contains a $R$-symmetric generator.
So one might guess the role played by the $R$ symmetry in $4d$ SUSY theory is crucial for
discriminating scale invariance from conformal invariance.

Following the intuition,
we address the connection between two symmetries in  the context of $4d$ $\mathcal{N}=1$ SUSY.
The first task we should solve is the realization of dilatation current of SUSY version.
This is tied to two elements.
At first, the SUSY version of dilatation current is actually to address the SUSY generalization of momentum-energy tensor $T_{\mu\nu}$ \cite{OS1, OS2}.
There are a few such multiplets known as supercurrent multiplets which admit the $T_{\mu\nu}$ as a component freedom.
These supercurrent multiplets
are classified into Ferrara-Zumino (FZ)-multiplet \cite{FZ}, $\mathcal{R}$- multiplet \cite{R} and
$\mathcal{S}$- multiplet \cite{1002.2228}, (see also \cite{CPS,1007.3092}).
The connection between scale and conformal invariance has been discussed in \cite{1102.2294} in terms of $\mathcal{R}$- multiplet,
which admits a conserved $R$ symmetry.
In this paper, we explore the FZ multiplet,
which conversely doesn't admit such conserved $R$ symmetry.

The other element which is crucial for our discussion is the ambiguity in the definition of $T_{\mu\nu}$.
As this is the main source for incorporating the scale and conformal invariance.
In what follows, we simply review the transformation of improvement due to the ambiguity in QFT,
and then take care of the improvement in its SUSY version.

The paper is organized as follows.
In section 2, inspired by the construction of dilatation current multiplet for $\mathcal{R}$-multiplet \cite{1102.2294},
we consider the SUSY version of dilatation current and virial current multiplet in the case of FZ multiplet.
In section 3, we use the consistent constraints of unitarity for scale-invariant SUSY theory
and closure of SUSY algebra as the main tool to explore the structure of the virial current multiplet.
We find that the the scale-invariant SUSY theory is also conformal when the real superfield in the dilatation current multiplet is conserved.
Otherwise, it is only scale-invariant, despite of the transformation of improvement.
Together with the claim on conditions for the equivalence between these two symmetries in $R$-symmetric case \cite{1102.2294},
we complete understanding their $4d$ SUSY version.

\section{Supercurrent and Dilatation Current Multiplet}
\subsection{$4D$ version in QFT}
Before we discuss the SUSY version of the dilatation current,
let us recall its definition in $4d$ quantum field theory (QFT).
Given a QFT with scaling invariance,
there exists a conserved current, i.e, the dilatation current $\Delta_{\mu}$,
which is found to be,
\begin{eqnarray}{\label{dc}}
\Delta_{\mu}=x^{\nu}T_{\mu\nu}+\mathcal{O}_{\mu}
\end{eqnarray}
Here $\mathcal{O}_{\mu}$ refers to the virial current that does not explicitly depend on the spacetime coordinates.
Scale invariance gives rise to an anomaly for the derivative of the virial current,
\begin{eqnarray}{\label{virial}}
 T=-\partial^{\mu}\mathcal{O}_{\mu}
\end{eqnarray}
which shows that the virial current must also be conserved if QFT with scale invariance is promoted to be conformal.
In other words, the anomaly involved the virial current in \eqref{virial} is a character for scale invariance against conformal invariance.
As illustrated by Polchinski in Ref. \cite{Polchinski},
scale-invariant QFT can be promoted to be a conformal one if and only if the virial current permits the structure as
\begin{eqnarray}{\label{decom}}
\mathcal{O}_{\mu}=j_{\mu}+\partial^{\nu}L_{\mu\nu},
\end{eqnarray}
with $L_{\mu\nu}$ is an anti-symmetric tensor and $j_{\mu}$ an conserved current, $\partial^{\mu}j_{\mu}=0$.

There are two important issues which determine whether virial current is allowed to have the structure as in \eqref{decom}.
The first issue is the ambiguity in the definition of the 4d energy-momentum tensor,
\begin{eqnarray}{\label{imp1}}
T_{\mu\nu}&\rightarrow&~T_{\mu\nu}+(\partial_{\mu}\partial_{\nu}-\eta_{\mu\nu}\partial^{2})\varphi
\end{eqnarray}
which transfers the ambiguity into the definition of virial current via \eqref{virial}.
The second issue involves in the SUSY version of \eqref{dc}.
Inspired by the connection between supersymmetric current $S_{\mu\alpha}$ and $T_{\mu\nu}$ from viewpoint of SUSY algebra,
they can be embedded into super-multiplets known as super-current multiplets.
In what follows, we choose the super-current multiplet which doesn't allow the $R$ symmetry, i.e, the FZ-multiplet $\mathcal{J}_{\mu}$
\footnote{
Supercurrent multiplets without conserved $R$ symmetry include $\mathcal{S}$- and FZ-multiplet.
Consider the fact that under certain limits \cite{1002.2228} the former reduces to either the later one or the $R$ multiplet,
we will study the FZ-multiplet.}.
We follow the conventions of Wess and Bagger \cite{Wess}
\footnote{The bi-spinor representation for vector field is
taken as,
\begin{eqnarray}\nonumber
J_{\alpha\dot{\alpha}}=-2\sigma^{\mu}_{\alpha\dot{\alpha}}J_{\mu},~~~~~~
and~~~~
J_{\mu}=\frac{1}{4}\bar{\sigma}_{\mu}^{\dot{\alpha}\alpha}J_{\alpha\dot{\alpha}}
\end{eqnarray}},
and present the explicit component expression for FZ-multiplet in the appendix A.
In the appendix, it is easy to see that the divergence of bottom component
$j_{\mu}=\partial^{\alpha\dot{\alpha}}\mathcal{S}_{\alpha\dot{\alpha}}=i\left(\bar{D}^{2}\bar{X}-D^{2}X\right)$,
from which this global current is not conserved except in the case of $\mathcal{R}$-multiplet.

\subsection{Dilatation Current Multiplet}

The SUSY version of \eqref{dc} can be realized by a multiplet $\mathbf{\Delta}_{\mu}$ defined as,
\begin{eqnarray}{\label{sdc}}
\mathbf{\Delta}_{\mu}=x^{\nu}\left\{-\frac{1}{8}\bar{\sigma}_{\mu}^{\dot{\alpha}\alpha}[D_{\alpha},
\bar{D}_{\dot{\alpha}}]\mathcal{\mathcal{J}}_{\nu}
+\frac{1}{16}\epsilon_{\nu\mu\rho\sigma}\left(\bar{\sigma}^{\dot{\alpha}\alpha}\right)^{\rho}\{D_{\alpha},\bar{D}_{\dot{\alpha}}\}\mathcal{J}^{\sigma}
+\frac{1}{4}\eta_{\mu\nu}(D^{2}X+\bar{D}^{2}\bar{X})\right\}+\mathcal{O}_{\mu}\nonumber\\
\end{eqnarray}
with $\Delta_{\mu}$ being the bottom component of real superfield $\mathbf{\Delta}_{\mu}$ in \eqref{dc},
from which one can verify \eqref{dc} by using the component expression in the appendix A.
The SUSY version of \eqref{virial} can be directly read from \eqref{sdc},
\begin{eqnarray}{\label{susyvirial}}
\partial^{\mu}\mathcal{O}_{\mu}=\frac{3}{16}(\bar{D}^{2}\bar{X}+D^{2}X)
\end{eqnarray}

In parrel to the previous discussion in the $4d$ version of QFT,
it is crucial to figure out the ambiguity of superfield $\mathcal{O}_{\mu}$ defined in \eqref{sdc}
before we proceed to discuss its structure.
The constraint on supercurrent $\mathcal{J}_{\mu}$,
$\bar{D}^{\dot{\alpha}}\mathcal{J}_{\alpha\dot{\alpha}}=D_{\alpha}X$ is not affected by an improvement as \cite{1002.2228},
\begin{eqnarray}{\label{imp3}}
\mathcal{J}_{\alpha\dot{\alpha}}&\rightarrow& \mathcal{J}_{\alpha\dot{\alpha}}-2i\sigma^{\mu}_{\alpha\dot{\alpha}}\partial_{\mu}(Y-\bar{Y})\nonumber\\
X&\rightarrow&~X-\frac{1}{2}\bar{D}^{2}\bar{Y}
\end{eqnarray}
where $Y$ a chiral superfield.
The improvement \eqref{imp3} shifts both the energy-momentum tensor and the supersymmetry current simultaneously as,
\begin{eqnarray}{\label{sandt}}
S_{\mu\alpha}&\rightarrow& S_{\mu\alpha}+2i\left(\sigma_{\mu\nu}\right)^{\beta}_{\alpha}\partial^{\nu}Y\mid_{\theta^{\beta}}\nonumber\\
T_{\mu\nu}&\rightarrow&~T_{\mu\nu}-(\partial_{\mu}\partial_{\nu}-\eta_{\mu\nu}\partial^{2})Re~Y\mid
\end{eqnarray}
Substituting the second improvement in \eqref{imp3} into \eqref{susyvirial} leads to
the SUSY version of the improvement transformations,
\begin{eqnarray}{\label{virialimp}}
\partial^{\mu}\mathcal{O}_{\mu}\rightarrow
\partial^{\mu}\mathcal{O}_{\mu}-\frac{3}{32}\left(D^{2}\bar{D}^{2}\bar{Y}+\bar{D}^{2}D^{2}Y\right)
\end{eqnarray}

Here are a few comments in order.
It is obvious that the transformations in \eqref{sandt} don't violate the conservations of supersymmetric current
$\partial^{\mu}S_{\mu\alpha}=0$ and energy-momentum tensor $\partial^{\mu}T_{\mu\nu}=0$.
Nevertheless, it modifies the trace part of energy-momentum tensor as $T\rightarrow T-3\partial^{2}(ReY\mid)$.
As we will discuss later,
this character potentially interpolates the scale invariant and conformal invariant theories.
That is to say, in a scale-invariant SUSY theory,
if there exists a well-defined $Y$ from UV to deep IR energy scale
such that trace part of energy-momentum tensor $T'=0$ by the improvement,
then this SUSY theory is \em{actually}\em~super-conformal.
Otherwise, a scale invariant SUSY theory is \em{exactly}\em~ allowed to exist,
unless it doesn't satisfy the examination of unitary constraints (see below for more discussions).

\section{Constraints on Virial-current Multiplet}
Now we proceed to uncover the structure of virial current superfield through some consistent checks from
SUSY algebra and unitarity of scale invariant theories.
The fact that the supercharge $Q_{\alpha}$ has scaling dimension $1/2$ implies that,
\begin{eqnarray}{\label{C1}}
[Q_{\alpha},\Delta]=-\frac{i}{2}Q_{\alpha}=\int d^{3}x [Q_{\alpha},\Delta_{0}]
\end{eqnarray}
By using the component expression for FZ-multiplet in appendix A,
one obtains,
\begin{eqnarray}{\label{C2}}
\int d^{3}x \left(\mathcal{O}_{0\alpha}-\frac{i}{2\sqrt{2}}(\sigma_{0}\bar{\psi})_{\alpha}-i(\sigma^{\nu}_{0})_{\alpha}^{\beta}S_{\mu\beta}-\frac{i}{2}S_{0\alpha}\right)=0
\end{eqnarray}
where $\psi=\frac{\sqrt{2}}{3}(\sigma^{\mu}\bar{S}_{\mu})$ in the case of FZ-multiplet,
and $Q_{\mu\alpha}\equiv[Q_{\alpha}, \mathcal{O}_{\mu}]$ as in Ref. \cite{1102.2294}.
Then \eqref{C2} gives rise to,
\begin{eqnarray}{\label{C3}}
\mathcal{O}_{\mu\alpha}=\frac{i}{3}\sigma_{\mu\alpha\dot{\alpha}}\bar{\sigma}^{\nu\dot{\alpha}\beta}S_{\nu\beta}
+(\sigma_{\mu}^{\nu})^{\beta\delta}\partial_{\nu}\gamma_{\beta\delta\alpha}
+(\bar{\sigma}_{\mu}^{\nu})^{\dot{\beta}\dot{\delta}}\partial_{\nu}\gamma_{\dot{\beta}\dot{\delta}\alpha}
\end{eqnarray}
where $\gamma_{\beta\delta\alpha}$ and $\gamma_{\dot{\beta}\dot{\delta}\alpha}$
are local and gauge invariant operators of dimension $5/2$.

In the case of conformal field theory, 
there exist well-known bounds on dimension of local and gauge invariant
operators \cite{Mack}.
Similar situation happens in the case of non-conformal fixed points \cite{0801.1140},
in terms of which operators $\gamma_{\beta\delta\alpha}$ and $\gamma_{\dot{\beta}\dot{\delta}\alpha}$
(and higher spin operators) are found to satisfy \cite{1102.2294},
\begin{eqnarray}{\label{C4}}
(\sigma_{\mu}^{\nu})^{\beta\delta}\partial_{\nu}\gamma_{\beta\delta\alpha}=0,~~~
(\bar{\sigma}_{\mu}^{\nu})^{\dot{\beta}\dot{\delta}}\partial_{\nu}\gamma_{\dot{\beta}\dot{\delta}\alpha}=0
\end{eqnarray}
Thus, one finds,
\begin{eqnarray}{\label{C5}}
\mathcal{O}_{\mu\alpha}=\frac{i}{3}\sigma_{\mu\alpha\dot{\alpha}}\bar{\sigma}^{\nu\dot{\alpha}\beta}S_{\nu\beta}
+(\sigma_{\mu}^{\nu})^{\alpha}_{\beta}\partial_{\nu}\gamma_{\beta}
\end{eqnarray}

To proceed,  we impose the closure of SUSY transformation to extract possible information on $\gamma_{\beta}$ in \eqref{C5} and its descents.
It is straightforward to impose the constraints,
\begin{eqnarray}{\label{susyt}}
(\eta^{\beta}\xi^{\alpha}-\xi^{\beta}\eta^{\alpha})\delta_{\beta}\delta_{\alpha}\mathcal{O}_{\mu}&=&0\nonumber\\
(\xi^{\alpha}\bar{\eta}_{\dot{\alpha}}\delta^{\dot{\alpha}}\delta_{\alpha}-\bar{\eta}_{\dot{\alpha}}\xi^{\alpha}\delta_{\alpha}\delta^{\dot{\alpha}})\mathcal{O}_{\mu}
&=&2i(\xi\sigma^{\nu}\bar{\eta})\partial_{\nu}\mathcal{O}_{\mu}\\
(\xi^{\alpha}\bar{\eta}_{\dot{\alpha}}\delta^{\dot{\alpha}}\delta_{\alpha}-\bar{\eta}_{\dot{\alpha}}\xi^{\alpha}\delta_{\alpha}\delta^{\dot{\alpha}})\gamma_{\beta}
&=&2i(\xi\sigma^{\nu}\bar{\eta})\partial_{\nu}\gamma_{\beta}\nonumber
\end{eqnarray}
which will give us some insights about the structure of $\gamma_{\beta}$.
In what follows, we follow a set of definitions of Ref.\cite{1102.2294},
\begin{eqnarray}{\label{C6}}
\delta_{\alpha}\gamma_{\beta}&=&i\epsilon_{\alpha\beta}\gamma-(\sigma^{\mu\nu})_{\alpha\beta}\gamma_{\mu\nu},\nonumber\\
\delta_{\dot{\alpha}}\gamma_{\beta}&=&(\sigma^{\mu})_{\beta\dot{\alpha}}\gamma_{\mu}
\end{eqnarray}
where $\gamma$, $\gamma_{\mu}$ and $\gamma_{\mu\nu}$ is gauge invariant scalar, vector and anti-symmetric tensor operator, respectively.

In terms of the SUSY transformation \eqref{A5},
 we obtain from the first constraint in \eqref{susyt},
\begin{eqnarray}{\label{C7}}
i\partial^{\nu}\gamma_{\nu\mu}+\frac{1}{3}\partial_{\mu}\gamma=0
\end{eqnarray}
Note that \eqref{C7} coincides with what has been found in the case of $\mathcal{R}$-multiplet.
Thus, as discussed in \cite{1102.2294},
we arrive at the conclusion that scalar $\gamma$ and tensor field $\gamma_{\mu\nu}$ both vanish.
In other words, $\gamma_{\alpha}$ is an anti-chiral superfield,
\begin{eqnarray}{\label{C8}}
D^{2}\mathcal{O}_{\mu}=\bar{D}^{2}\mathcal{O}_{\mu}=0
\end{eqnarray}

Evaluating the second constraint in \eqref{susyt},
one derives that,
\begin{eqnarray}{\label{C9}}
\partial_{\nu}\mathcal{O}_{\mu}=-\frac{2}{3}\eta_{\nu\mu}T-\frac{2}{3}\epsilon_{\mu\nu\rho\sigma}\partial^{\rho}j^{\sigma}
-\frac{1}{4}\partial_{\nu}(\gamma_{\mu}+\bar{\gamma}_{\mu})+\frac{1}{4}\eta_{\nu\mu}
\partial^{\rho}(\gamma_{\rho}+\bar{\gamma}_{\rho})
-\frac{i}{4}\varepsilon_{\sigma\rho\mu\nu}\partial^{\sigma}(\gamma^{\rho}-\bar{\gamma}^{\rho})\nonumber\\
\end{eqnarray}
where we have used the anti-commutators of supercharges and supercurrent.
From \eqref{C9} one finds the divergence of virial current,
\begin{eqnarray}{\label{C10}}
\partial^{\mu}\mathcal{O}_{\mu}=-\frac{1}{20}\partial^{\mu}(\gamma_{\mu}+\bar{\gamma}_{\mu})
\end{eqnarray}
in term of the relation $T=-\partial^{\mu}\mathcal{O}_{\mu}$.
Introduce superfield $\Gamma_{\alpha}$ which accommodate $\gamma_{\alpha}$ as the bottom component,
$\Gamma_{\alpha}=\gamma_{\alpha}+\cdots$,
we can write \eqref{C10} in the superfield expression
\begin{eqnarray}{\label{C11}}
\mathcal{O}_{\mu}&=&\frac{1}{40}\bar{\sigma}_{\mu}^{\dot{\alpha}\alpha}
\left(\bar{D}_{\dot{\alpha}}\Gamma_{\alpha}-D_{\alpha}\bar{\Gamma}_{\dot{\alpha}}\right)+J_{\mu}
\end{eqnarray}

Following the fact that the anti-symmetric part involved $\gamma_{\mu}$ in \eqref{C9} doesn't contribute to $\mathcal{O}_{\mu}$ in \eqref{C11},
one can accommodate this part as,
\begin{eqnarray}{\label{antis}}
U_{\mu}=-i(\gamma_{\mu}-\bar{\gamma}_{\mu})+\mathcal{\hat{O}}_{\mu}
\end{eqnarray}
where $U_{\mu}$ and $\mathcal{\hat{O}}_{\mu}$ is the vector freedom of
real superfield $U$ \footnote{For illustration, superfield $U$ that can be decomposed into a chiral and its anti-chiral superfield,
can achieve the null contribution from $U_{\mu}\sim \partial_{\mu}(A-A^{*})$. }
and a primary operator $\mathcal{\hat{O}}$.
Using the last constraint in \eqref{susyt}, one finds
\begin{eqnarray}{\label{C12}}
\Gamma_{\alpha}= \frac{i}{2}D_{\alpha}U +  \frac{1}{2}\mathcal{\hat{O}}_{\alpha}
\end{eqnarray}
from which \eqref{C11} can be rewritten as
\begin{eqnarray}{\label{C13}}
\mathcal{O}_{\mu}&=&-\frac{1}{2}\partial_{\mu}U+\frac{1}{8}\bar{\sigma}_{\mu}^{\dot{\alpha}\alpha}
\left[D_{\alpha},\bar{D}_{\dot{\alpha}}\right]\mathcal{\hat{O}}
\end{eqnarray}
where we have made a scaling of $\mathcal{O}_{\mu}$.

In summary, the virial current multiplet in a scale-invariant SUSY theory satisfies
\begin{eqnarray}{\label{C14}}
0&=&D^{2}\mathcal{O}_{\mu}=\bar{D}^{2}\mathcal{O}_{\mu}\nonumber\\
\mathcal{O}_{\mu}&=&-\frac{1}{2}~\partial_{\mu}U+\frac{1}{8}\bar{\sigma}_{\mu}^{\dot{\alpha}\alpha}
\left[D_{\alpha},\bar{D}_{\dot{\alpha}}\right]\mathcal{\hat{O}}
\end{eqnarray}

\section{Scale Invariance vs Conformal Invariance}
According to \eqref{virialimp} the scale-invariant SUSY theory can be improved to be conformal-invariant  if and only if,
\begin{eqnarray}{\label{con1}}
\partial^{\mu}\mathcal{O}_{\mu}=\{D^{2},\bar{D}^{2}\}\hat{Y}
\end{eqnarray}
with $\hat{Y}=Y+\bar{Y}$ a real superfield.
Impose the first constraint of \eqref{C14} on its second one,
one immediately finds that
\begin{eqnarray}{\label{con2}}
\partial^{\mu}\mathcal{O}_{\mu}=-\frac{1}{4}~\Box~U
\end{eqnarray}
If $U$ is conserved, the scale-invariant SUSY theory is actually conformal-invariant.
Conversely, it is only scale-invariant,
which is not affected by the improvement \eqref{con1} as we explain below.

Compare \eqref{con2} with \eqref{con1},
one gets the intuition that scale-invariant SUSY with non-conserved $U$ can be improved to be conformal-invariant
if $U$ satisfies,
\begin{eqnarray}{\label{con4}}
\left(\Box~-\tilde{c}~\{D^{2},\bar{D}^{2}\}\right)U=0
\end{eqnarray}
with an adjustable real coefficient $\tilde{c}$.
Also the improvement suggests that $U$ is proportional to $\hat{Y}$.
This means $\hat{Y}$ should also satisfy \eqref{con4} and $D^{2}\hat{Y}=0$ simultaneously.
Substituting the later into \eqref{con4} leads to,
\begin{eqnarray}{\label{con5}}
\Box~\hat{Y}=0~~~~~\Leftrightarrow ~~~~~\Box~U=0
\end{eqnarray}
In conclusion, when the virial current multiplet $\mathcal{O}_{\mu}$, defined by the scale-invariant SUSY theory,
doesn't contain a conserved $U$, the theory can not be improved to be conformal.
Conversely, when such an $U$ is conserved, the scale-invariant SUSY theory must also be conformal,
despite of the transition of improvement.
Together with the claim on the equivalence between these two symmetries in $R$-symmetric case \cite{1102.2294},
we complete understanding their $4d$ SUSY version.

To illustrate the roles played by $R$ symmetry,
let us use the SUSY Wess-Zumino model for example.
Given the Kahler potential $K(\Phi^{i},\bar{\Phi}^{i})$ and superpotential $W(\Phi_{i})$ for chiral superfields $\Phi_{i}$,
the FZ-multiplet and $X$ superfield is given by ,
\begin{eqnarray}{\label{con6}}
\mathcal{J}_{\alpha\dot{\alpha}}&=& 2g_{i\bar{i}}\left(D\Phi^{i}\right)\left(\bar{D}\bar{\Phi}^{\bar{i}}\right)
-\frac{2}{3}[D_{\alpha},D_{\dot{\alpha}}]K,\nonumber\\
X&=&4W-\frac{1}{3}\bar{D}^{2}K
\end{eqnarray}
If there is $R$ symmetry in SUSY Wess-Zumino models,
 $X$ can be written as
the specific form \cite{1002.2228},
\begin{eqnarray}
X=\bar{D}^{2}\left(\frac{1}{2}\sum_{i}R_{i}\Phi^{i}\partial_{i}K-\frac{1}{3}K\right)=-\frac{1}{2}\bar{D}^{2}\tilde{U},
\end{eqnarray}
It is crucial to note that $U$ is identifies as $\tilde{U}$ that is indeed decomposed of a chiral and its anti-chiral part.
Such $U$ which is constrained by $D^{2}U=\bar{D}^{2}U=0$ as in \eqref{C14} trivially satisfies \eqref{con5}.
Superficially, the super-conformality in this type of $R$-symmetric Wess-Zumino model
is restored in terms of the transition of improvement.
From the viewpoint of virial current multiplet,
this improvement is actually irrelevant.  \\

~~~~~~~~~~~~~~~~~~~~~~~~~~~~~~~~~~~~~~~~
$\bf{Acknowledgement}$\\
This work is supported in part by the Doctoral Fund of Ministry of Education of China (No. 20110191120045).

\appendix
\section{Communicators}
The component expression for $\mathcal{S}_{\mu}$ that satisfies the constraint
$\bar{D}^{\dot{\alpha}}\mathcal{J}_{\alpha\dot{\alpha}}=D_{\alpha}X+\chi_{\alpha}$ is given by \cite{1002.2228},
\begin{eqnarray}{\label{A1}}
\mathcal{J}_{\mu}&=& j^{(S)}_{\mu}+\theta^{\alpha}(S_{\mu\alpha}-\frac{1}{\sqrt{2}}\sigma_{\mu}\bar{\psi})
+\bar{\theta}(\bar{S}_{\mu}+\frac{1}{\sqrt{2}}\bar{\sigma}_{\mu}\psi)+\frac{i}{2}\theta^{2}\partial_{\mu}\phi^{\dag}
-\frac{i}{2}\bar{\theta}^{2}\partial_{\mu}\phi\nonumber\\
&+&(\theta\sigma^{\nu}\bar{\theta})\left(2T_{\nu\mu}-\eta_{\nu\mu}Z+\frac{1}{2}\epsilon_{\mu\nu\rho\sigma}\left(F^{(S)\rho\sigma}+\partial^{\rho}j^{(S)\sigma}\right)\right)\\
&+&\theta^{2}\left(\frac{i}{2}\partial_{\rho}S_{\mu}\sigma^{\rho}-\frac{i}{2\sqrt{2}}\partial_{\rho}\bar{\psi}\bar{\sigma}^{\rho}\sigma_{\mu}\right)\bar{\theta}
+\bar{\theta}^{2}\theta\left(-\frac{i}{2}\sigma^{\rho}\partial_{\rho}S_{\mu}+\frac{i}{2\sqrt{2}}\sigma_{\mu}\bar{\sigma}^{\rho}\partial_{\rho}\psi\right)\nonumber\\
&+&\theta^{2}\bar{\theta}^{2}\left(\frac{1}{2}\partial_{\mu}\partial^{\nu}j_{\nu}^{(S)}-\frac{1}{4}\partial^{2}j^{(S)}_{\mu}\right)\nonumber
\end{eqnarray}
with
\begin{eqnarray}{\label{A2}}
X&=&\phi+\sqrt{2}\theta\psi+\theta^{2}\left(Z+i\partial^{\rho}j^{(R)}_{\rho}\right)\nonumber\\
\chi_{\alpha}&=&-i\lambda^{(S)}_{\alpha}+\left(D\delta_{\alpha}^{\beta}
-2i(\sigma^{\rho}\bar{\sigma}^{\sigma})_{\alpha}^{~\beta}~F^{(S)}_{\rho\sigma}\right)\theta_{\beta}
+\theta^{2}\sigma_{\nu\alpha\dot{\alpha}}\partial_{\nu}\bar{\lambda}^{(S)\dot{\alpha}}
\end{eqnarray}
The component fields also satisfy two extra constraints,
\begin{eqnarray}{\label{A3}}
D=-4T^{\mu}_{\mu}+6Z,~~~~~~\lambda_{\alpha}^{(S)}=-2i\sigma^{\mu}\bar{S}_{\mu}+3i\sqrt{2}\psi.
\end{eqnarray}
According to \eqref{A2}, one can derive the SUSY transformation of supercurrent $S_{\mu\alpha}$,
\begin{eqnarray}{\label{A4}}
\delta_{\dot{\beta}}S_{\mu\alpha}&=&\sigma^{\nu}_{\alpha\dot{\beta}}
\left(2T_{\nu\mu}-i\eta_{\nu\mu}\partial^{\rho}j^{(S)}_{\rho}+i\partial_{\nu}j^{(S)}_{\mu}
-\frac{1}{2}\epsilon_{\nu\mu\rho\sigma}F^{(S)\rho\sigma}
-\frac{1}{2}\epsilon_{\nu\mu\rho\sigma}\partial^{\rho}j^{(S)\sigma}\right)\nonumber\\
\delta_{\beta}S_{\mu\alpha}&=&-2\varepsilon_{\lambda\beta}(\sigma_{\mu\rho})^{\lambda}_{\alpha}\partial^{\rho}\phi^{*}
\end{eqnarray}
as well as their conjugators,
\begin{eqnarray}{\label{A5}}
\delta_{\beta}\bar{S}_{\mu\dot{\alpha}}&=&\sigma^{\nu}_{\beta\dot{\alpha}}
\left(2T_{\nu\mu}+i\eta_{\nu\mu}\partial^{\rho}j^{(S)}_{\rho}-i\partial_{\nu}j^{(S)}_{\mu}
-\frac{1}{2}\epsilon_{\nu\mu\rho\sigma}F^{(S)\rho\sigma}
-\frac{1}{2}\epsilon_{\nu\mu\rho\sigma}\partial^{\rho}j^{(S)\sigma}\right)\nonumber\\
\delta_{\dot{\beta}}\bar{S}_{\mu\dot{\alpha}}&=&-2\varepsilon_{\dot{\lambda}\dot{\beta}}(\bar{\sigma}_{\mu\rho})^{\dot{\lambda}}_{\dot{\alpha}}\partial^{\rho}\phi.
\end{eqnarray}

We can obtain the corresponding relationes for the case of  FZ multiplet by taking the limit $\chi_{\alpha}=0$.
The anti-commutators relationes \eqref{A4} and \eqref{A5} modify as,
\begin{eqnarray}{\label{A6}}
\delta_{\dot{\beta}}S_{\mu\alpha}&=&\sigma^{\nu}_{\alpha\dot{\beta}}
\left(2T_{\nu\mu}-i\eta_{\nu\mu}\partial^{\rho}j_{\rho}+i\partial_{\nu}j_{\mu}
-\frac{1}{2}\epsilon_{\nu\mu\rho\sigma}\partial^{\rho}j^{\sigma}\right)\nonumber\\
\delta_{\beta}S_{\mu\alpha}&=&-2\varepsilon_{\lambda\beta}(\sigma_{\mu\rho})^{\lambda}_{\alpha}\partial^{\rho}\phi^{*}
\end{eqnarray}
and
\begin{eqnarray}{\label{A7}}
\delta_{\beta}\bar{S}_{\mu\dot{\alpha}}&=&\sigma^{\nu}_{\beta\dot{\alpha}}
\left(2T_{\nu\mu}+i\eta_{\nu\mu}\partial^{\rho}j_{\rho}-i\partial_{\nu}j_{\mu}
-\frac{1}{2}\epsilon_{\nu\mu\rho\sigma}\partial^{\rho}j^{\sigma}\right)\nonumber\\
\delta_{\dot{\beta}}\bar{S}_{\mu\dot{\alpha}}&=&-2\varepsilon_{\dot{\lambda}\dot{\beta}}(\bar{\sigma}_{\mu\rho})^{\dot{\lambda}}_{\dot{\alpha}}\partial^{\rho}\phi.
\end{eqnarray}
Here we have used the $D=0$ and $\lambda^{(S)}_{\alpha}=0$ in \eqref{A3} to cancel the dependence on $\psi$ and $Z$ fields,
from which the conservation of $\partial^{\mu}S_{\mu\alpha}=0$ is consistent with these anti-commutators.


\begin{thebibliography} {99}

\bibitem{Zamolodchikov}
A. B. Zamolodchikov, ``Irreversibility of the Flux of the Renormalization Group in a
2D Field Theory ,''JETP Lett. {\bf43}, 730 (1986) [Pisma Zh. Eksp. Teor. Fiz. 43, 565
(1986)].

\bibitem{Polchinski}
J. Polchinski, ``Scale and Conformal Invariance in Quantum Field Theory ,''
Nucl. Phys. B {\bf303}, 226 (1988).

\bibitem{Callan}
C. G. Callan, S. R. Coleman and R. Jackiw, ``New improved energy - momentum
tensor ,''Annals Phys. {\bf59}, 42 (1970).


\bibitem{0910.1087}
D. Dorigoni and V. S. Rychkov, ``Scale Invariance + Unitarity $\Rightarrow$ Conformal Invari-
ance? '' [arXiv:0910.1087].

\bibitem{0904.2715}
D. Gaiotto, ``N=2 dualities,''
[arXiv:0904.2715].

\bibitem{9511154}
P. C. Argyres, M. Ronen Plesser, N. Seiberg and E. Witten, ``New N=2 Superconfor-
mal Field Theories in Four Dimensions ,'' Nucl. Phys. B 461, 71 (1996) [arXiv:hep-
th/9511154].

\bibitem{9411149}
N. Seiberg, `` Electric - magnetic duality in supersymmetric nonAbelian gauge theo-
ries ,'' Nucl. Phys. B{\bf435}, 129 (1995), [arXiv:hep-th/9411149].

\bibitem{Mack}
G. Mack,
`` All Unitary Ray Representations Of The Conformal Group SU(2,2) With
Positive Energy,''
Commun. Math. Phys. {\bf55}, 1 (1977).

\bibitem{1102.2294}
I. Antoniadis, M.Buican,
``On R-symmetric Fixed Points and Superconformality,''
[arXiv:1102.2294].



\bibitem{OS1}
V. Ogievetsky and E. Sokatchev ``Supercurrent,''
Sov. J. Nucl. Phys{\bf28},~423~(1978),
Yad. Fiz{\bf28}, 825~(1978).

\bibitem{OS2}
V. Ogievetsky and E. Sokatchev, ``On vector superfield generated by supercurrent,'' Nucl.
Phys. B {\bf124}, 309 (1977).

\bibitem{FZ}
S. Ferrara and B. Zumino, ``Transformation Properties Of The Supercurrent,'' Nucl.
Phys. B {\bf87}, 207 (1975).

\bibitem{R}
K. S. Stelle and P. C. West,`` Minimal Auxiliary Fields For Supergravity,'' Phys. Lett.
B{\bf74}, 330 (1978).

\bibitem{1002.2228}
Z. Komargodski, N. Seiberg,
``Comments on Supercurrent Multiplets, Supersymmetric Field Theories and Supergravity,''
JHEP {\bf07},~017~(2010),
[arXiv:1002.2228].

\bibitem{CPS}
T. E. Clark, O. Piguet and K. Sibold,
``Supercurrents, Renormaliztion and Anomalies,''
Nucl. Phys. B{\bf143},~445~(1978).

\bibitem{1007.3092}
S. Zheng and J. Huang,``Variant supercurrent and Linearized Supergravity,''
Class. Quant. Grav. {\bf28},~075012~(2011).
arXiv: [arXiv:1007.3092].

\bibitem{Wess}
J. Wess and J. Bagger, ``Supersymmetry and supergravity,'' Princeton Univ.
Pr. (1992).

\bibitem{0801.1140}
B. Grinstein, K. A. Intriligator and I. Z. Rothstein, ``Comments on Unparticles ,''
Phys. Lett. B {\bf662}, 367 (2008) [arXiv:0801.1140].











\end{thebibliography}
\end{document}